\documentclass[letterpaper]{article}
\usepackage[top=3cm,bottom=3cm,left=3cm,right=3cm,marginparwidth=1.75cm]{geometry}
\usepackage{listings}
\usepackage{xcolor}
\usepackage{multirow}
\usepackage{dcolumn}
\usepackage[english]{babel}
\usepackage[utf8x]{inputenc}
\usepackage[T1]{fontenc}
\usepackage{palatino}
\def\ga{\gamma}

\def\De{\Delta}

\def\nue{\nu_e}
\def\numu{\nu_\mu}

\def\nuebar{\bar\nu_e}
\def\numubar{\bar\nu_\mu}

\def\lsim{\mathrel{\rlap{\lower4pt\hbox{\hskip1pt$\sim$}}
    \raise1pt\hbox{$<$}}}
\def\gsim{\mathrel{\rlap{\lower4pt\hbox{\hskip1pt$\sim$}}
    \raise1pt\hbox{$>$}}}

\newcommand{\beq}{\begin{eqnarray}}
\newcommand{\eeq}{\end{eqnarray}}
\def\to{\rightarrow}

\usepackage{amsmath}
\usepackage{afterpage}
\usepackage{graphicx, subcaption}
\usepackage[colorinlistoftodos]{todonotes}
\usepackage[colorlinks=true, linkcolor=blue, citecolor=blue]{hyperref}

\begin{document}

\vspace*{1.2cm}
\thispagestyle{empty}
\begin{center}

{\LARGE \bf MiniBooNE Neutrino Oscillation Search Results and Predicted Background Events}

\par\vspace*{7mm}\par

{

\bigskip

\large \bf Teppei Katori for the MiniBooNE collaboration}

\bigskip

{\large \bf  e-mail: teppei.katori@kcl.ac.uk}

\bigskip

{King's College London, Strand, WC2R 2LS London, UK}

\bigskip

{\it Presented at the 3rd World Summit on Exploring the Dark Side of the Universe \\Guadeloupe Islands, March 9-13 2020}

\end{center}

\begin{abstract}
  In this short review, we discuss the 2020 MiniBooNE electron neutrino appearance oscillation results with a special attention on background predictions relevant to the MiniBooNE oscillation results and other (anti)electron neutrino appearance search experiments.  
\end{abstract}
  
\section{MiniBooNE 2020 Oscillation results}
\label{S:intro}

\begin{figure}
\begin{center}
  \vspace{-15pt}
  \includegraphics[width=5in]{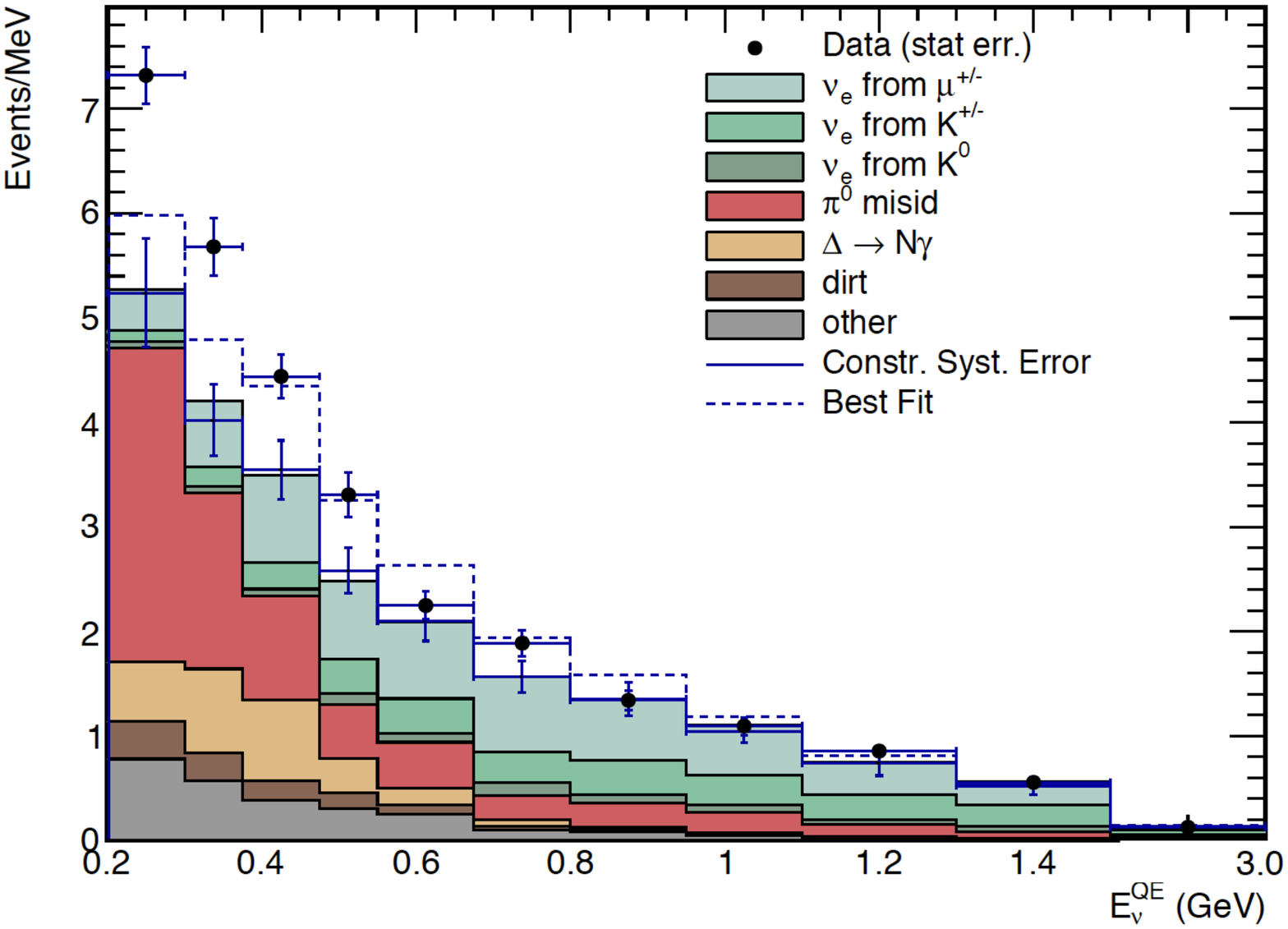}
  \vspace{-25pt}
\end{center}
\caption{
MiniBooNE data-MC comparison for the $\nue$ oscillation candidate event distribution~\cite{Aguilar-Arevalo:2020nvw}. This is a function of reconstructed electron neutrino energy under the QE assumption ($E_\nu^{QE}$). 
}
\label{fig:Fig1}
\end{figure}

MiniBooNE was a short-baseline neutrino oscillation experiment at Fermilab. A $\numu$ ($\numubar$) dominant beam was created by the Booster neutrino beamline (BNB)~\cite{AguilarArevalo:2008yp}. A mineral-oil-based spherical Cherenkov detector~\cite{AguilarArevalo:2008qa}, located 541~m away from the target, was used to search for single-isolated electron-like signals produced by charged-current (CC) interactions of $\nue$ ($\nuebar$) neutrinos. Data excesses of signals over backgrounds have been reported~\cite{Aguilar-Arevalo:2018gpe}.  Data taking was stopped in 2019, and in 2020 we presented the results from the 17-year full data set~\cite{Aguilar-Arevalo:2020nvw}. Fig.~\ref{fig:Fig1} shows the final result of $\nue$ candidate spectrum as a function of $E_\nu^{QE}$, the reconstructed neutrino energy under the quasi-elastic (QE) assumption~\cite{AguilarArevalo:2007ab} which assumes two-body kinematics and a target nucleon at rest. The data excess can be interpreted as a signal of neutrino oscillations at a $\sim 1$~eV mass scale, or as the presence of sterile neutrinos around the $1$~eV mass scale. MiniBooNE signals are statistically the strongest signals of the so-called short-baseline anomalies~\cite{Diaz:2019fwt} which all suggest a $\sim 1$~eV sterile neutrinos, and there are world-wide programs to search for such a neutrino candidate.

The data exhibit an excess of events over simulated backgrounds in the lower energy region. Here, 6 main backgrounds are explicitly shown. In brief, beam-origin backgrounds ($\nue$ from $\mu^{\pm}$, $\nue$ from $K^{\pm}$, and  $\nue$ from $K^0$) are intrinsic $\nue$ backgrounds, and they tend to spread over wide energy region. This feature makes it difficult to explain the excess by a mis-modelling of the beam without exotic effects~\cite{Giunti:2007xv}. On the other hand, other backgrounds ($\pi^\circ$ misid, $\De\to N\ga$, dirt) are mis-identification (misID) backgrounds, mainly photons that are misidentified as electrons from $\nue$CCQE interactions. These backgrounds have similar shapes as the excess, and we discuss these backgrounds further in this short note.

\section{$\pi^\circ$ misid}

Every experiments searching for $\nue$($\nuebar$) appearance signals, neutral current single $\pi^\circ$ (NC1$\pi^\circ$) production is a serious background channel. Although $\pi^\circ$ normally decays to two gamma rays, boosted $\pi^\circ$ can decay to one gamma ray, or one gamma ray is undetected, then the final state is a single gamma which is resemble to a single-isolated electron-like event by Cherenkov detectors. The prediction of such background is difficult, because the prediction of neutrino produced single pion is very difficult. This problem comes from two parts. First, we need to predict production rates of pions from the primary process correctly. Such calculation relies on advanced nuclear models of baryonic resonances and non-resonant meson productions from neutrino interactions. Models are tuned and tested with electron scattering data~\cite{Hernandez:2013jka}, however, axial form factors need to be tuned from  low statistics neutrino scattering experiments~\cite{Wilkinson:2014yfa}.  Confusions increase if the energy goes higher and higher resonances~\cite{Nakamura:2015rta} (beyond the $\De$ resonance) and transition region to DIS (so-called shallow inelastic scattering, SIS~\cite{Andreopoulos:2019gvw,SajjadAthar:2020nvy,Alvarez-Ruso:2020ezu}) are non-negligible. Second part is the final state interactions (FSIs). Hadrons experience effects from the nuclear environment, and these modify both kinematics and types of hadrons leaving the target nuclei. Simulations of FSIs for hadrons are also difficult, and thus correct simulation of photon background is very challenging. To overcome these problems, MiniBooNE utilizes an internal constraint by measuring $\pi^\circ$ events in MiniBooNE~\cite{AguilarArevalo:2008xs}. 
These data are used to tune the simulation of $\pi^\circ$ spectrum. Then, the photon background from $\pi^\circ$s is simulated by performing decays of data-tuned $\pi^\circ$s.

\begin{figure}
\begin{center}
  \vspace{-15pt}
  \includegraphics[width=5in]{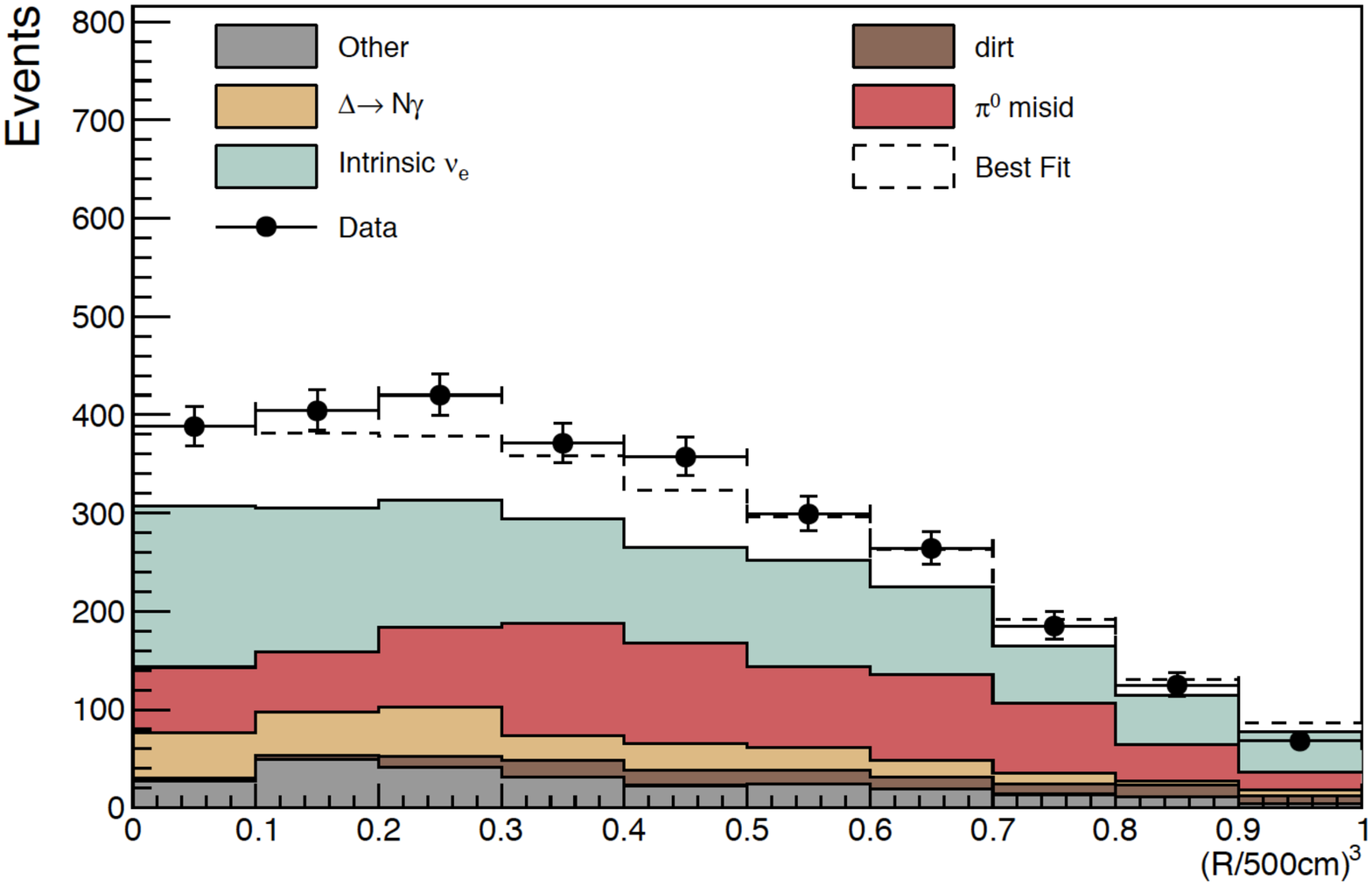}
  \vspace{-25pt}
\end{center}
\caption{Oscillation candidate data-MC comparison with function of (R/500cm)$^3$~\cite{Aguilar-Arevalo:2020nvw}, note 500~cm is the fiducial volume of the MiniBooNE detector. $\pi^\circ$ background makes a peak slightly off-centre, which do not agree with the shape of the data excess. 
}
\label{fig:Fig2}
\end{figure}

Fig.~\ref{fig:Fig2} shows a coordinate distribution of data and simulated background events with function of cube of normalized radial distance R from the center of the detector. Note, fiducial volume of the MiniBooNE detector is a 500 cm spherical region of the mineral oil volume. Because of the geometric effect, the photon background from $\pi^\circ$s makes a characteristic shape and the peak is not around the centre. Absence of this feature in the data suggest NC1$\pi^\circ$ cannot explain the data excess.

\section{$\De\to N\ga$}

\begin{figure}
\begin{center}
  \vspace{-15pt}
  \includegraphics[width=6in]{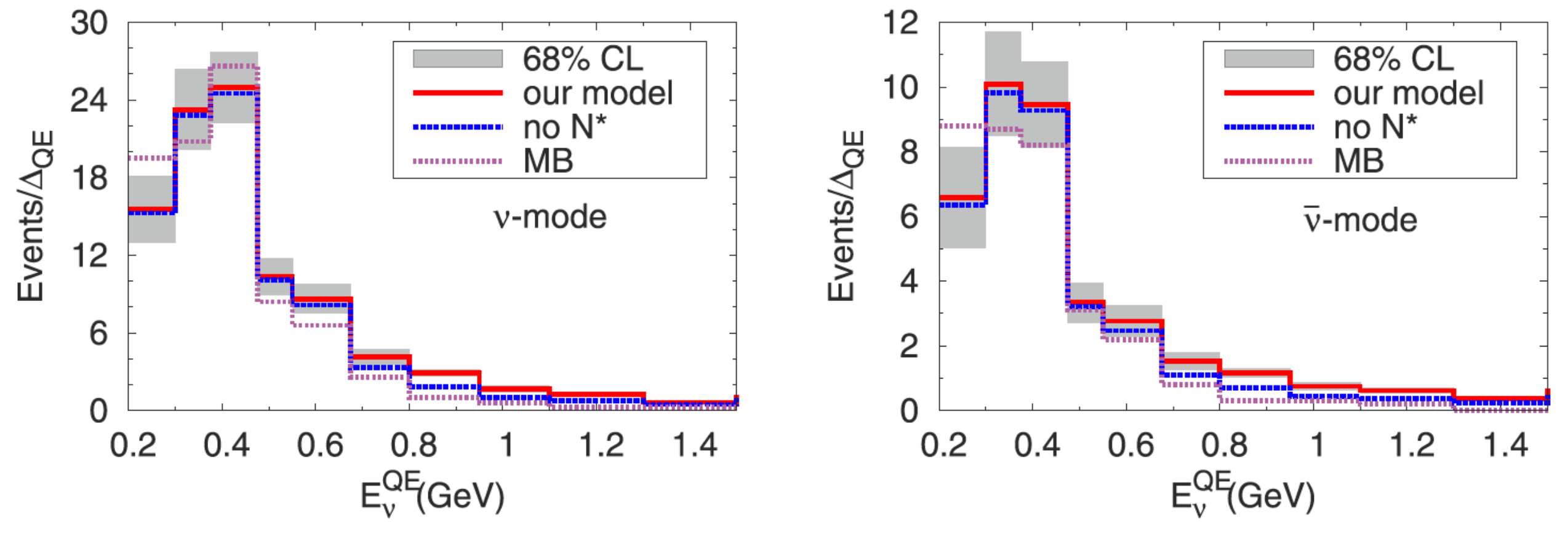}
  \vspace{-25pt}
\end{center}
\caption{
  A comparisons of MiniBooNE prediction of $\De\to N\ga$ backgrounds and the Valencia NC1$\ga$  model~\cite{Wang:2014nat}. Agreements are within the theoretical errors except the lowest bins.
  Similar agreement is obtained from other modern NC1$\ga$ models~\cite{Hill:2010zy,Zhang:2012xn}. 
}
\label{fig:Fig3}
\end{figure}

A high-energy single photon can be made by a radiative decay of a baryonic resonance. In MiniBooNE, such NC single gamma (NC1$\ga$) channel is not simulated by Monte Carlo, instead, the prediction is made from measured $\pi^\circ$ rate. In brief, measured $\pi^\circ$ rate is used to extrapolate the $\De$-resonance rate after correcting the FSIs, then branching ratio is applied to predict the single gamma ray background via the radiative $\De$-decay. As Fig.~\ref{fig:Fig3} shows, the MiniBooNE prediction agreed with the predictions from the latest nuclear calculations~\cite{Wang:2014nat}. Similar results are obtained from other advanced NC1$\ga$ models~\cite{Hill:2010zy,Zhang:2012xn}. These theoretical calculations give us a confidence that we predict this background correctly. However, the radiative $\De$-decay has not been experimentally confirmed, and only limits are available~\cite{Kullenberg:2011rd,Abe:2019cer}. Therefore, it might be possible to explain all excesses by the radiative $\De$-decay if the theoretical prediction is wrong around 300\%. Furthermore, it is also possible that exotic processes (mainly $Z'$-decay) contribute similar signals~\cite{Ballett:2016opr,Bertuzzo:2018itn,Arguelles:2018mtc,Jordan:2018qiy}. These possibilities are testable soon by high-resolution detectors such as the MicroBooNE experiment (liquid argon time projection chamber)~\cite{Acciarri:2016smi} and the NINJA experiment (emulsion cloud chamber)~\cite{Hiramoto:2020gup}.

\section{dirt}

High-energy photons generated outside of the detector can come inside of the detector without showering in the veto region. This type of background, called dirt events in MiniBooNE, is challenging to simulate since materials and geometry outside of the detector are not completely modeled, and prediction is not reliable. To estimate dirt background, we use a data-driven correction. Events outside of the fiducial volume are measured to find a distribution of dirt events with a function of R, the distance from the detector center. This gives a better estimation of the dirt background inside of the fiducial volume. 

\begin{figure}
\begin{center}
  \vspace{-15pt}
  \includegraphics[width=4in]{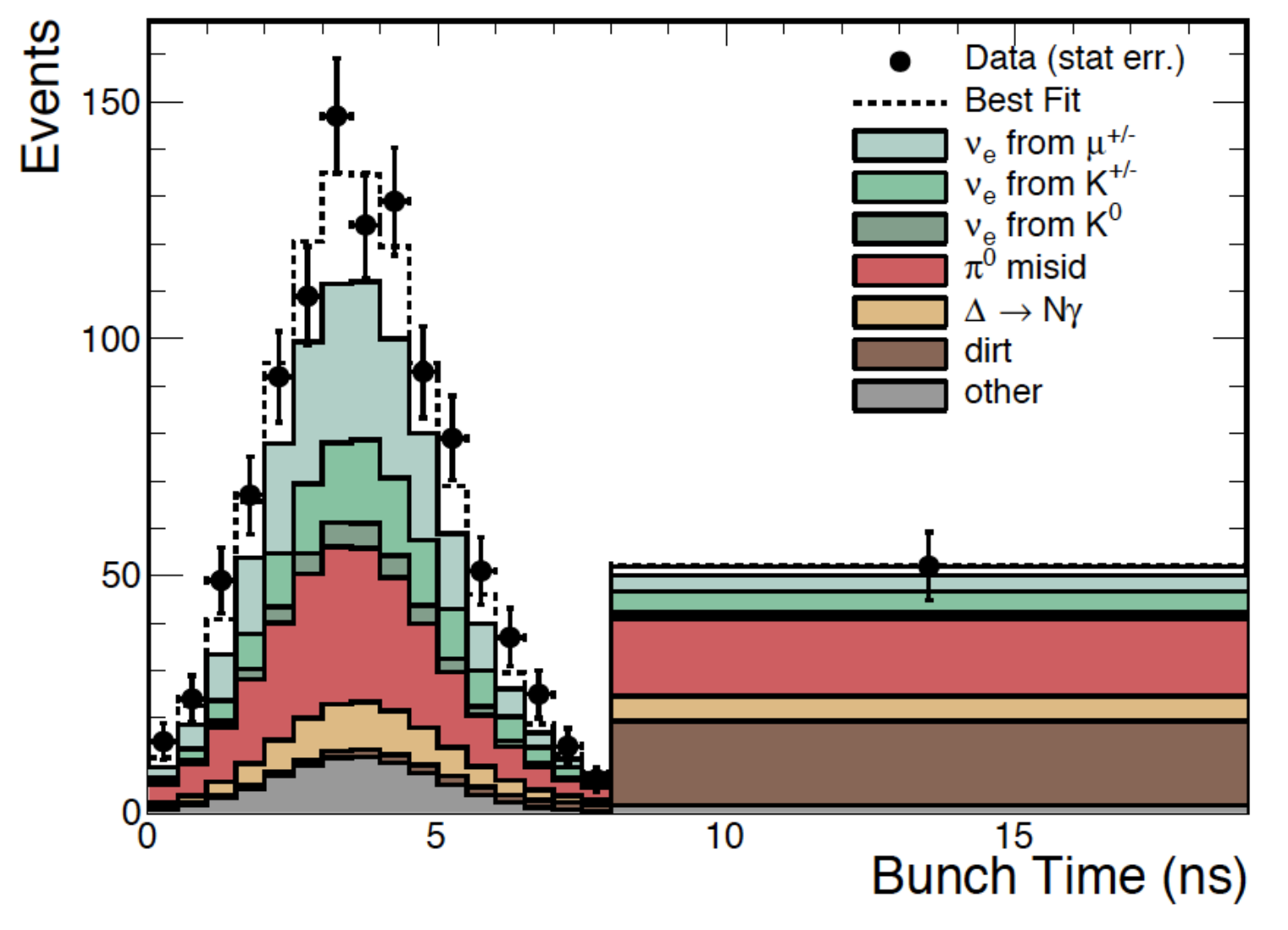}
  \vspace{-25pt}
\end{center}
\caption{
  Oscillation candidate event time distribution in the beam bunch timing~\cite{Aguilar-Arevalo:2020nvw}. The bunch time is $\sim$2~ns with $\sim$19~ns separation to repeat 82 times. The dirt event is out of the phase of bunch time and it makes a peak outside of the bunch timing. Both intrinsic backgrounds and misID backgrounds show slight delays compared with oscillation signal events from $\nue$CCQE interactions. 
}
\label{fig:Fig4}
\end{figure}

We also imply the timing information to further check this background. Among the 84 BNB buckets, $\sim$83 buckets are usually filled, where 2~ns bunches are separated with 19~ns. Fig.~\ref{fig:Fig4} shows the bunch timing of $\nue$ candidate events with background simulations. This precise timing structure is used for the search of the beam produced boosted dark matter~\cite{Aguilar-Arevalo:2017mqx,Aguilar-Arevalo:2018wea}. Here, the timing data is used to look for sterile neutrinos. First, dirt background is out of phase and we can confirm our prediction of this background is correct from the data-MC agreement outside of the bunch timing.  Second, beam-origin intrinsic backgrounds show a slight delay ($\sim$1~ns) due to heavier parents (kaon decay background) or additional processes (muon decay background). Third, misID photon backgrounds also show a slight delay ($\sim$1~ns) due to additional showering process compared with electron oscillation candidate signals from $\nue$CCQE interaction. Note, timing information is not currently used to select $\nue$($\nuebar$) oscillation candidates.

\section{Nucleon correlations}

\begin{figure}
\begin{center}
  \vspace{-15pt}
  \includegraphics[width=6in]{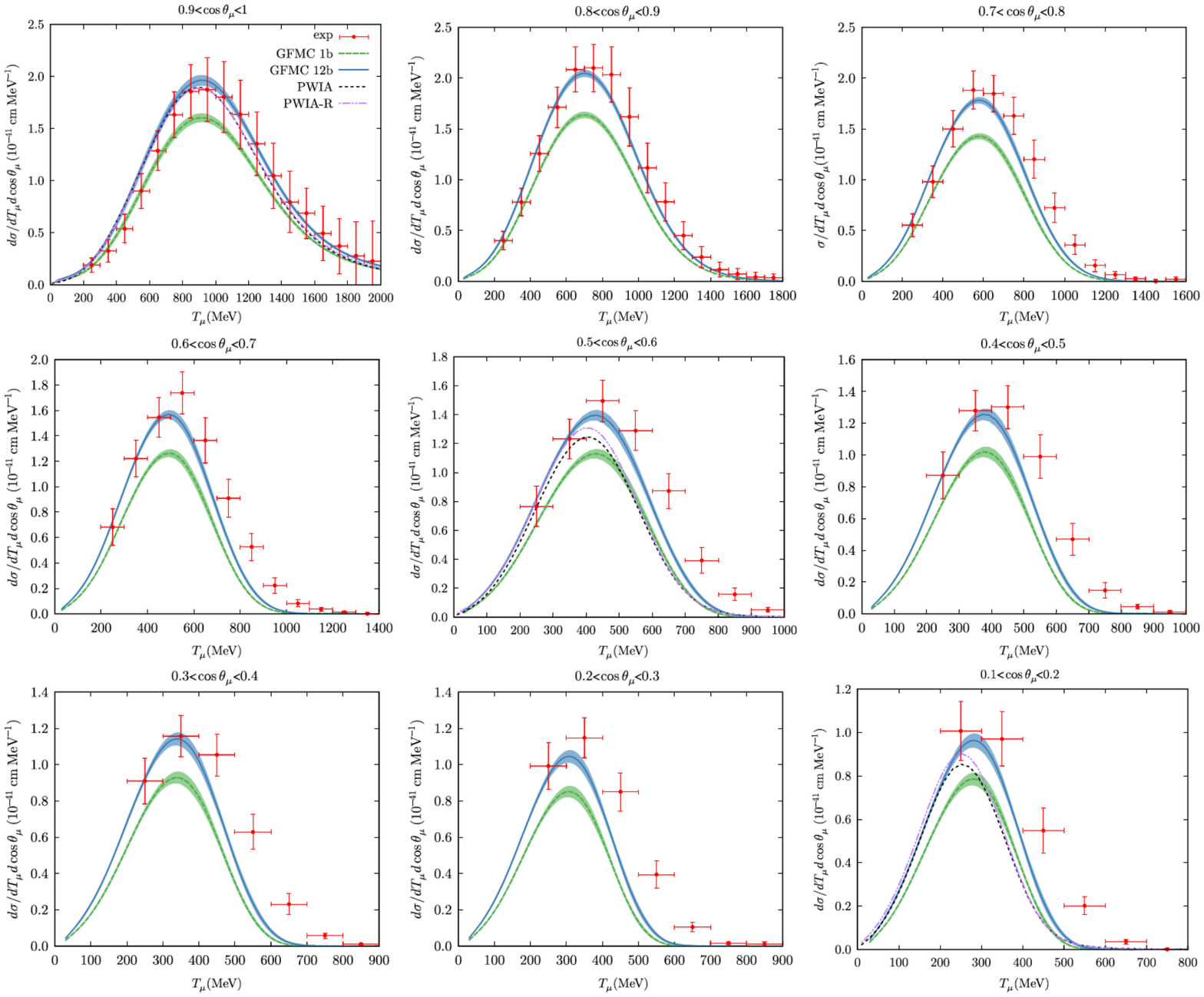}
  \vspace{-25pt}
\end{center}
\caption{
  MiniBooNE $\numu$CCQE double differential cross-section data~\cite{AguilarArevalo:2010zc} are compared with {\it ab initio} quantum Monte Carlo prediction~\cite{Lovato:2020kba}. In the theory, all nuclear potentials are included, and five response functions are calculated in wide range of momentum transfer, from 100 to 700 MeV.
}
\label{fig:Fig5}
\end{figure}

Neutrino interaction physics around 1-10 GeV has extremely rich structures~\cite{Alvarez-Ruso:2017oui}, and future long-baseline oscillation experiments, such as DUNE~\cite{Abi:2020wmh} and Hyper-Kamiokande (HyperK)~\cite{Abe:2018uyc}, are likely to be systematically limited including neutrino interaction errors. Among them, the discovery of the role of nucleon correlations~\cite{Martini:2009uj,Nieves:2011pp} in neutrino interactions attract many interests and actively studied by many experiments. Neutrinos often interact with more than one nucleon, or correlated nucleon pair. This changes outgoing particle kinematics, and also this enhances the cross-section more than incoherence sum of all nucleon contributions. This additional channel, often called MEC (meson exchange current) or 2p2h (two-particle 2-hole), are introduced in modern neutrino interaction generators~\cite{Sobczyk:2012ms,Lalakulich:2012ac,Schwehr:2016pvn,Abe:2017vif}. Fig.~\ref{fig:Fig5} shows the data-theory comparison of MiniBooNE flux-integrated $\numu$CCQE double differential cross-section data~\cite{AguilarArevalo:2010zc} and {\it ab initio} quantum Monte Carlo (QMC) calculation~\cite{Lovato:2020kba}. This precise nuclear theory includes both two-nucleon and three-nucleon potentials, and cross-section predictions capture the detailed features of MiniBooNE high-statistics $\numu$ cross-section data. Since MiniBooNE does not include this channel in the simulation and effect of cross-section enhancement is tuned through CCQE channel, it was argued that data excess could be related to the neutrino energy mis-reconstruction~\cite{Martini:2012fa,Nieves:2012yz}. This idea is interesting, because energy mis-reconstruction would shift the energy spectrum to the lower energy, namely the data excess would be higher energy region more consistent with sterile neutrino oscillation hypothesis. So far, detailed study does not support this~\cite{Ericson:2016yjn}, but the jury is still out. Current knowledge of nucleon correlations in neutrino physics is very limited, and experiments require to tune this channel a lot to improve data-MC agreements within their simulations~\cite{Abe:2017vif,Ruterbories:2018gub,Acero:2020eit}. This may have some implications on a mild tension of oscillation results between T2K~\cite{Abe:2019vii} and NOvA~\cite{Kelly:2020fkv}. 

\section{$\nue$($\nuebar$) cross-sections}

Although neutrino pion production and 2p2h models are tested many times with $\numu$($\numubar$)CC data, $\nue$($\nuebar$) appearance oscillation experiments need these models to provide correct predictions for $\nue$($\nuebar$)CC interactions. This is difficult to test, mainly because neutrino beams including the BNB~\cite{AguilarArevalo:2008yp},  J-PARC neutrino beamline~\cite{Abe:2012av}, and NuMI~\cite{Adamson:2015dkw} do not produce large $\nue$($\nuebar$) flux in the region where experiments expect $\numu\to\nue$ ($\numubar\to\nuebar$) oscillations. 
To take account this error, we introduced a neutrino-energy dependent $\nue/\numu$ ratio error which blows up exponentially at low energy. It is only few \% at 400 MeV but over 30\% error at 200~MeV. There are only 5 low statistics data available for $\nue$ cross-section~\cite{Blietschau:1977mu,Abe:2014agb,Wolcott:2015hda,Acciarri:2020lhp,Abe:2020vot} in entire particle physics history. Thus, we have unavoidable systematic errors on $\nue$($\nuebar$) interactions, and this is an issue for all  $\nue$($\nuebar$) appearance experiments including MiniBooNE, T2K, NOvA, and future experiments such as DUNE and HyperK. 

\section{Conclusions}

In this brief review, we go through some of key developments of our background study. The statistical significance of the excess is very high, but currently there is no convincing explanation of this through any existing background channels.
%Future developements include further study of kinematics, nucleon correlatino channels, and timing information. 

\section*{Acknowledgements}

I thank Bill Louis and Rex Tayloe for their careful checks of this article. I thank the hospitality of the  organizers for hosting the participants during this conference even though the pandemic made  the organization extremely difficult.

\bibliographystyle{JHEP.bst} 
\bibliography{references.bib}

%\printindex

\end{document}